\documentstyle[11pt]{article}
\def\rlx{\relax\leavevmode}
\def\inbar{\vrule height1.5ex width.4pt depth0pt}
\def\IC{\rlx\hbox{\,$\inbar\kern-.3em{\rm C}$}}
\begin{document}

\begin{titlepage}

\begin{flushright}
NTZ  37/98, YERPHI 1528(1)-99 
\end{flushright}

\vspace{2cm}

\begin{center}
  {\large\bf  Conserved currents of the three-reggeon interaction.}
\\[1cm]

  {\bf D.R.~Karakhanyan} \\[5mm]
  {\it  Yerevan Physics Institute,\\
      375036, Yerevan,  Armenia} \\[1cm]

  {\bf  R. ~Kirschner} \\

 {\it Naturwissenschaftlich-Theoretisches Zentrum und \\ 
Institut f\"ur Theoretische Physik,\\
   Universit\"at Leipzig, \\
  D-04109, Leipzig, Germany}

\end{center}
\vspace{2cm}
\abstract{ We consider an extension of  Lipatov's conjecture about
the deep relation between amplitudes in the high-energy limit of QCD
and XXX Heisenberg chains with non-compact spins.}

\vspace{5cm}
{\sl Talk given by D.R.K. at the INTAS-RFBR School,Como, Italy,  September 98}
\end{titlepage}

\section{Introduction}
In Regge theory the high-energy asymptotics of the hadron-hadron
scattering amplitudes is determined by singularities of partial waves
in the complex angular momentum  plane.
It was observed \cite{1} that the regularity of quantum mechanics, which
relates high-energy scattering amplitudes to the singularities of the partial
waves in the complex angular momentum plane is valid for quantum field
theory as well.
Namely, in Regge kinematics
\begin{equation}
\label{1}
s\gg-t\sim M^2
\end{equation}
where $M$ is a characteristic hadronic mass scale, the hadron-hadronic
scattering amplitude $A(s,t)$
\begin{equation}
\label{2}
A(s,t)=is\int\limits^{\epsilon+i\infty}_{\epsilon-i\infty}\frac{dJ}{2\pi i}
\left(\frac{s}{M^2}\right)^J f(J,t)
\end{equation}
is governed by singularities of the partial waves $f(J,t)$, by Regge
poles and Regge cuts. Among the Regge singularities there is one
with vacuum quantum
numbers, the so-called Pomeron, which provides the dominant contribution
to the
scattering amplitude. The bootstrap conjecture has been
proposed, according to
which all particle-like excitations  correspond to some Regge
singularity
and are related to each other via unitarity of the S-matrix and sum
rules.

However the programme of building the axiomatic quantum field theory from
assumptions of only unitarity and analyticity of the S-matrix failed,
because Regge
theory itself does not allow to calculate the positions of these
singularities.
Now,  QCD as a theory of strong interaction is called  to describe
the Regge behavior of the scattering amplitudes \cite{2}.
V.N. Gribov proposed the conjecture, that reggeons form  new
collective excitations
and QCD in the high-energy limit can be replaced by an effective reggeon
theory
\cite{3}. This has been confirmed in series of works, initiated by
L.N.Lipatov
\cite{4}. It was shown that in leading logarithmic approximation (LLA),
which
is the natural approximation in the Regge limit of QCD,  $A(s,t)$ can be
expressed as a sum of Feynman
diagrams describing the multiple  exchange of reggeized gluons in the
$t$-channel. The
perturbative expressions for corresponding Feynman diagrams  including
large
logarithmic factors $\alpha^nlog^ms$ (m=n,n-1,...) have to be resummed
to
all orders in $\alpha_s$, because bare gluons and quarks are not a good
approximation in the Regge limit.
The leading contribution $(m=n)$ comes from ladder diagrams,
corresponding to exchange of $n$-reggeons in $t$-channel. Being built 
from an infinite number of perturbative gluons, the reggeons carry the
quantum numbers of the gluon and become a new collective
excitation in the Regge limit (1.1). It is well known that the leading
logarithmic
approximation results in an asymptotics, which violates the Froissart
bound.
Unitarity is restored by taking into account sub-leading contributions
as
well. In the generalised leading logarithmic approximation (GLLA) some
minimal
set of non-leading terms is included to restore unitarity \cite{7}. The
interaction of the reggeons is determined by LLA.

In LLA the dominant contribution to the  partonic scattering amplitude
comes from
the soft gluons and this leads to the gluon reggezation property. It can
be shown
that infrared divergences are cancelled for colourless external states due
gauge invariance.

Of course the distribution of partons inside the hadron are  described
by the
nonperturbative wave function of the hadron. The non-perturbative
effects can
be taken into account in the approach of constructing the high-energy
effective
action \cite{5}. However, the perturbative investigation of
high-energy QCD
makes sense as a first approximation to begin with. In particular
results
obtained in the BFKL pomeron approximation are in a good agreement with
experimental data of semi-hard processes and especially of
deep-inelastic
scattering at small $x$. Then, some information about non-perturbative
corrections can be extracted from analyzing the behaviour of the
perturbative series in the infrared region \cite{6}.

\section{High-energy QCD as an integrable model}
As mentioned above, the high-energy asymptotics of the scattering
amplitude
in leading logarithmic approximation is determined by the contribution of
diagrams, describing two reggeon exchange in the $t$-channel. It is a result of
summing up an infinite number of the famous ladder diagrams \cite{4} and
corresponds to the exchange of the Pomeron. Contributions of diagrams
with three
(Odderon) and more reggeized gluons can be considered as  higher
corrections.
Pomeron contribution corresponds to elastical  scattering. Diagrams
with Odderon exchange describe processes with the exchange of negative
charge parity.

The dominating contribution in LLA comes from the multi-Regge
kinematics:
\begin{eqnarray}
\label{3}
s=(p_A+p_B)^2\approx 2p_A\cdot p_B\\
\nonumber
s_i=(k_i+k_{i+1})^2\approx 2k_i\cdot k_{i-1}\\
\nonumber
i=1,..n+1,\quad k_0=p_A,\quad k_i=q_{i+1}-q_i,\quad k_{n+1}=p_B\\
\nonumber
s\gg s_i\gg\mid q_i\mid^2\\
\nonumber
s_1s_2\cdot\cdot\cdot s_{n+1}=s\prod_{i=1}^n(-k_{\bot i}^2)\\
\nonumber
k_1\cdot p_A\ll k_2\cdot p_A\ll...\ll k_n\cdot p_A\\
\nonumber
k_1\cdot p_B\gg k_2\cdot p_B\gg...\gg k_n\cdot p_B
\nonumber
\end{eqnarray}
where $k_\bot$ is defined by the Sudakov decomposition
$$k^\mu={k\cdot p_A\over\displaystyle p_A\cdot
p_B}p_B^\mu+{k\cdotp_B\over\displaystyle p_A\cdot p_B}
p_A^\mu+k_\bot^\mu$$
Owing to this decomposition the
scattering amplitudes in the Regge limit exibit the remarkable
separation
of the longitudinal and transverse directions with respect to the plane,
spanned by the momenta of the  initial particles.

In the generalized leading logarithmic approximation
the interaction between reggeons is elastic and pairwise.  We shall
restrict ourselves to the case where  the number of
reggeons in the $t$-channel $N$ is conserved. For a given $N$ the
reggeon Green function $ f_{\{i_k\}} $ satisfies the Bethe-Salpeter-like
equation \cite{7}
\begin{equation}
\label{4}
\omega f_{\{{i_{k}}\}}=f^{(0)}_{\{i_k\}}
+\sum_{i<j}{\cal{H}}^{(i,j)}_{\{i_k\},\{j_k\}} f_{\{j_k\}}
\end{equation}

The set $\{i_k\}=(i_1,..i_r)$ labels the reggeons, $i=G$ stands for
gluon and $i=F$ or $i=\bar F$ stand for reggeized quarks of
corresponding
helicity. The partial wave $f$ carries also an index $\alpha_i$,
labelling the
gauge group representation of the corresponding reggeons and depends
in their transverse momentum $k_{\bot i}$ or their impact parameter
$x_i$.

The $r$-reggeon contribution to the partial wave is obtained by
contracting
the r-reggeon Green function with the parton distribution functions of
the scattered particles. The angular momentum is
\begin{equation}
\label{5}
J=1+\omega-{r_f\over 2}
\end{equation}
where $r_f$ is the number of exchanged fermions.
The pairwise interaction of the reggeons is described by the Hamiltonian
\begin{eqnarray}
\label{6}
{\cal H}^{(i,j)}f_{i_1,..i,..i_r}={g^2\over(2\pi)^3}\int dk^{\prime}_i
d{k^{\prime}}_j\delta(k_i+k_j-{k^{\prime}}_j-k^{\prime}_j)[
(T_i\otimes T_j)_{\cal H}{\cal H}_{i,j}f_{i_1,..i,..j,..i_r}+\\
(T_i\otimes T_j)_{\cal G}{\cal G}_{i,j}f_{i_1,..i,..j,..i_r}]\nonumber
\end{eqnarray}
The first term in the square bracket corresponds to the interaction via
an $s$-channel gluon:
\begin{equation}
\label{7}
(T_i\otimes T_j)_{{\cal H}_{\{\alpha_k,{\alpha\prime}_k\}}}=\prod^r
_{k\ne i,j}\delta_{\alpha_k,{\alpha\prime}_k}
(T^a)_{\alpha_i{\alpha\prime}_i}
(T^a)_{\alpha_j{\alpha\prime}_j}
\end{equation}
and the second one corresponds to the interaction via an $s$-channel
fermion
\begin{equation}
\label{8}
(T_i\otimes T_j)_{{\cal G}_{\{\alpha_k,\alpha\prime_k\}}}=\prod^r
_{k\ne i,j}\delta_{\alpha_k,{\alpha_k}\prime}
(T^{{\alpha_i}\prime})_{\alpha_i\alpha}
(T^{\alpha_j})_{\alpha\alpha\prime_j}
\end{equation}
with the sum over the fermion colour states $\alpha$. The overall group
state in the $t$-channel has to be the gauge singlet.

While the longitudinal part of the scattering amplitude is extracted as
a kinematical factor, the transverse dynamical part can be described by
simple Feynmann rules, corresponding to the multi-regge effective
action \cite{5}.
These graphical rules allow the simple derivation of the interaction
kernels, entering the equation (\ref{6}). As operators in impact space
these kernels take the following form
\begin{eqnarray}
\label{9}
{\cal H}_{GG}=H_G+H_G^\ast\\
\nonumber
{\cal H}^{\omega}_{F\bar F}=H^\omega_F+P_{12}H^{\ast(\omega)}_FP_{12}\\
\nonumber
{\cal H}_{FF}=H_G+H_F^\ast\\
\nonumber
{\cal H}_{FG}=H_G+ {\cal P}_{12}H_F^\ast {\cal P}_{12}  \\
\nonumber
{\cal G}_{FG}=(x_{12}^\ast\partial_2^\ast)^{-1}
\nonumber
\end{eqnarray}
where
\begin{eqnarray}
\label{10}
H_G=-2\psi(1)+
\partial_1^{-1}\log x_{12}\partial_1+
\partial_2^{-1}\log x_{12}\partial_2+\log \partial_1\partial_2\\
\nonumber
H_F^\omega=-2\psi(1)+
\partial_1^{-1+\omega/2}\log x_{12}\partial_1^{1-\omega/2}+
\partial_2^{-\omega/2}\log x_{12}\partial_2^{\omega/2}+
\log \partial_1\partial_2
\end{eqnarray}
and $\psi(1)=-\gamma$ is the Euler number.  ${\cal P}_{12} $ represents
the operator permuting the reggeons 1 and 2. 

L.N.Lipatov \cite{8} applied the quantum inverse scattering method
to solve
equations for wave functions of compound states of $n$ reggeized gluons.
The eigenvalue problem related to the operators (\ref{10}) arises,
because the
position of the singularities in $\omega$ of the $t$-channel partial
wave $f_{\{i_k\}}$  is determined by their eigenvalues.

Lipatov proposed to diagonalize the $N$ reggeon problem by establishing
the  correspondence between the operators
(\ref{10}) and the $XXX$ Heisenberg model. He has noticed that
the operator $H_G$ has two equivalent representations:
\begin{eqnarray}
\label{11}
H_G=\partial_1^{-1}\log x_{12}\partial_1+
\partial_2^{-1}\log x_{12}\partial_2+\log\partial_1\partial_2-2\psi(1)=
\\ \nonumber
x_{12}\log\partial_1\partial_2 x_{12}^{-1}+2\log x_{12}-2\psi(1)
\end{eqnarray}
and therefore the transposed operator can be represented in two ways as
follows:
\begin{equation}
\label{12}
(H_G)^T=\partial_1\log x_{12}\partial_1^{-1}+
\partial_2\log x_{12}\partial_2^{-1}+\log\partial_1\partial_2-2\psi(1)=
\partial_1\partial_2 H_G(\partial_1\partial_2)^{-1}.
\end{equation}
On the  other hand we have
\begin{equation}
\label{13}
(H_G)^T=x_{12}\log\partial_1\partial_2x_{12}^{-1}+2\log x_{12}-2\psi(1)=
x_{12}^{-2}H_Gx_{12}^2
\end{equation}
Then we can deduce
\begin{equation}
\label{14}
[H_G;{\cal C}^{00} ] = 0, \ \ \ {\cal C}^{00} =
x_{12}^2\partial_1\partial_2
\end{equation}
This equation expresses the fact of conformal invariance of $H_G$.
Indeed, this operator coincides with Casimir operator of $SL(2)$ of
zero conformal weights ${\cal C}^{00}$. Then it is reasonable to denote
$H_G$ as $H^{00}$.

The hamiltonian
\begin{eqnarray}
\label{15}
H_{123}=H_{12}^{00}+H_{23}^{00}+H_{31}^{00}=\\\nonumber
\partial_1^{-1}\log x_{31}x_{12}\partial_1+
\partial_2^{-1}\log x_{12}x_{23}\partial_2+
\partial_3^{-1}\log x_{23}x_{31}\partial_3+
2\log\partial_1\partial_2\partial_3-6\psi(1)=\\
x_{12}\log\partial_1\partial_2x_{12}^{-1}+
x_{23}\log\partial_2\partial_3x_{23}^{-1}+
x_{31}\log\partial_1\partial_3x_{31}^{-1}
+2\log x_{12}x_{23}x_{31}+6\psi(1)\nonumber
\end{eqnarray}
corresponds
to the exchange
of\\
three reggeized
gluons (Fig.1).
\qquad
\parbox{7cm}{
\unitlength=1mm
\linethickness{0.2mm}
\begin{picture}(30,20)
\put(5,5){\circle{3}}
\put(15,22){\circle{3}}
\put(25,5){\circle{3}}
\put(6.5,5){\line(1,0){17}}
\put(5,6.5){\line(3,5){8.6}}
\put(25,6.5){\line(-3,5){8.6}}
\put(2,7){\shortstack{1}}
\put(14,24){\shortstack{2}}
\put(26,7){\shortstack{3}}
\put(10,0){\shortstack{Fig.1}}
\end{picture}
}
\quad

The transposition gives
\begin{equation}
\label{16}
(H_{123})^T=\partial_1\partial_2\partial_3H_{123}(\partial_1\partial_2
\partial_3)^{-1}=(x_{12}x_{23}x_{31})^{-1}H_{123}x_{12}x_{23}x_{31},
\end{equation}
i.e. it commutes with the operator $A_3$:
\begin{equation}
\label{17}
[H_G;A_3]=0,\quad\quad A_3=x_{12}x_{23}x_{31}\partial_1\partial_2\partial_3
\end{equation}
Notice, that operator $A_3$ is the commutator of partial Casimir
operators of the chain links:
\begin{equation}
\label{18}
A_3=[{\cal C}^{00}_{12},{\cal C}^{00}_{31}]=[{\cal C}^{00}_{23},{\cal
C}^{00}_{12}]= [{\cal C}^{00}_{31},{\cal C}^{00}_{23}]
\end{equation}
Therefore eq. (\ref{17}) is a consequence of (\ref{18}) and
the Jacobi identity. It is easy to check, that the $N$-reggeon
hamiltonian

\begin{equation}
\label{19}
H_N=\sum_{i=1}^NH_{i,i+1},\quad\quad H_{N,N+1}\equiv H_{N,1}
\end{equation}
commutes with
\begin{equation}
\label{20}
A_N=(\prod_{i=1}^{N-1} x_{i,i+1})(\prod_{j=1}^N\partial_j)
\end{equation}
It can be checked also, that $H_N$ commutes with ${\cal C}_N$,
\begin{equation}
\label{21}
{\cal C}_N=\sum_{1\leq i<j\leq N}x_{ij}^2\partial_i\partial_j.
\end{equation}
The eigenvalue problem:
\begin{equation}
\label{22}
\tilde A\Psi_\Delta(x_1,x_2,x_3)=\tilde a\Psi_\Delta(x_1,x_2,x_3),
\qquad\tilde{\cal C}
\Psi_\Delta(x_1,x_2,x_3)=\Delta(1-\Delta)\Psi_\Delta(x_1,x_2,x_3),
\end{equation}
can be considered instead of the original eigenvalue problem:
\begin{equation}
\label{23}
\tilde H_{123}\Psi_\Delta(x_1,x_2,x_3)=\tilde E_{123}\Psi_\Delta(x_1,x_2,x_3),
\end{equation}
which looks much more complicated.

The  equation (\ref{21}) expresses the conformal invariance of the $N$
reggeon
system and suggests the connection with the isotropic $sl(2)$ Heisenberg
model with $N$ cites and cyclic boundary conditions, because ${\cal C}_N$
can be identified with Casimir operator of this symmetry algebra.

Lipatov has related to each !site of the Heisenberg chain the following
Lax operator
\begin{equation}
\label{24}
L_i=\left(\matrix{\lambda+S_i^0&S_i^-\cr-S_i^+&\lambda-S_i^0\cr}\right)
\end{equation}
where the $SL(2)$ spin operators $S_i^a$
\begin{equation}
\label{25}
[S^0_i;S^\pm_j]=\pm\delta_{ij}S_i^\pm,\quad\quad
[S_i^+;S_j^-]=-2\delta_{ij}S_i^0
\end{equation}
are represented as differential operators
\begin{equation}
\label{26}
S_i^+=x_i^2\partial_i+2\Delta_ix_i,\quad\quad
S_i^0=x_i\partial_i+\Delta_i,\quad\quad S_i^-=\partial_i
\end{equation}
and $\lambda$ is the spectral parameter.
The Lax operator (\ref{21}) satisfies the Yang-Baxter
equation. Therefore the  trace of the monodromy matrix
\begin{equation}
\label{27}
T(\lambda)=\prod_{i=1}^NL_i(\lambda)
\end{equation}
is the generating function for the set of $N$ mutually commuting
differential operators $Q_k$:
\begin{equation}
\label{28}
Q_k=\sum_{i_1<i_2<...<i_k}x_{i_1i_2}x_{i_2i_3}...x_{i_ki_1}\partial_{i_1}
\partial_{i_2}...\partial_{i_k}
\end{equation}
Now it is plausible enough that all  operators $Q_k$ commute with the
Hamiltonian and the latter can be represented as a function of $Q_k$.
Explicite
calculations  for low $N$ cases and additional mathematical arguments
\cite{9} cofirm this.
Moreover, the relation between the high-energy QCD kernels and the $XXX$
Heisenberg
spin chains turns out to be much  deeper. Namely, this connection
can be extended
to the case where fermions are incorporated. The ideas of conformal
symmetry
in Regge asymptotics \cite{10} have been developed in application to the
fermion exchange in \cite{11}.

\section{Review of integrable chains}

Let us review the main facts of the theory of integrable systems,
concerning  the XXX Heisenberg magnet. In our review we shall follow
 Sklyanin's work \cite{12}. The phenomenon of the integrability of
quantum systems can be understood by means of their relation to
the linear ones via separation of variables. Namely, the quantum system is
 integrable, if its non-linear equations of motion can be represented as
the zero-curvature conditions of some integrable linear system \cite{13}.
Physically this means, that the interaction of  such systems reduces to
the elastic scattering and the only  result of it consists in the
exchange of quantum numbers (momenta etc.) of the scattered particles.
Accordingly, the $S$-matrix of the theory is factorized into the
product of blocks, corresponding to
$2\rightarrow 2$ scattering and also
$1\rightarrow 1$ in the presence of a boundary \cite{14}.

A set of (annihilation)
operators $Z_a(\lambda)$ satisfying the Zamolodchikov algebra:
has been proposed for an algebraic description of the factorizable
scattering.

\begin{equation}
\label{29}
Z_a(\lambda)Z_b(\mu)=S_{ab,cd}(\lambda-\mu)Z_d(\mu)Z_c(\lambda)
\end{equation}
where $S$ is $n^2\times n^2$ matrix.
The consistency condition of this system, which follows from
the associativity property of the triple product
$Z_{a_1}(\lambda_1)Z_{a_2}(\lambda_2)Z_{a_3}(\lambda_3)$,
that is the Yang-Baxter equation for the $S$-matrix:
\begin{eqnarray}
\label{30}
S_{a_ja_k}(\lambda_j-\lambda_k)
S_{a_ja_l}(\lambda_j-\lambda_l)
S_{a_ka_l}(\lambda_k-\lambda_l)=\\\nonumber
S_{a_ka_l}(\lambda_k-\lambda_l)
S_{a_ja_l}(\lambda_j-\lambda_l)
S_{a_ja_k}(\lambda_j-\lambda_k).\nonumber
\end{eqnarray}
Extending this algebra by adding n conjugated (creation) operators
$Z_a^\dagger(\mu)$, one gets the Zamolodchikov - Faddeev algebra:
\begin{equation}
\label{31}
Z_a(\lambda)Z_b^\dagger(\mu)=\delta_{ab}\delta(\lambda-\mu)+
Z_c^\dagger(\mu)\hat S_{ac,bd}(\lambda-\mu)Z_d(\lambda),
\end{equation}
or in matrix notations:
\begin{eqnarray}
\label{32}
A(\lambda)\otimes A(\mu)\equiv A_1(\lambda)A_2(\mu)=
S_{12}(\lambda-\mu)A_2(\mu)A_1(\lambda),\\\nonumber
A_1^\dagger(\lambda)A_2^\dagger(\mu)=A_2^\dagger(\mu)A_1^\dagger(\lambda)
S_{21}^\dagger(\mu-\lambda),\\\nonumber
A_1(\lambda)\otimes A_1^\dagger(\mu)=I_1\delta(\lambda-\mu)+
A_2^\dagger(\mu)\hat S_{12}(\mu-\lambda)A_2(\lambda),\nonumber
\end{eqnarray}
where $A(\lambda)$ and $A^\dagger(\mu)$ are the column
$(Z_1(\lambda),...,Z_n(\lambda))^t$ and the row
$(Z_1^\dagger(\mu),...,Z_n^\dagger(\mu)$
correspondingly, subscripts refer to the corresponding isotopic spaces
$\IC^n\otimes\IC^n\equiv V_1V_2$ and $S_{21}={\cal P}S_{12}{\cal P}$,
$\hat S_{12}={\cal P}S_{12}$, ${\cal P}$ is the permutation operator in
$\IC^n\otimes\IC^n$.
The complete scattering matrix $S(\{\lambda_k\})$ of the $M$ particle is
factorized then into the ordered product of $M(M-1)/2$ two-particle
$S$-matrices (\ref{29}). For example the $S$-matrix of the $j$-th
particle on the other
$M-1$ particles is given by $t(\lambda_j;\{\lambda_m\})$, i.e. the  
particular value of the transfer matrix for $\lambda=\lambda_j$:
\begin{equation}
\label{33}
t(\lambda;\{\lambda_m\})=tr_aT(\lambda;\{\lambda_m\})\equiv tr_a\prod_kS_
{ak}(\lambda-\lambda_k),
\end{equation}

The trace in this expression is taken over the auxiliary space $V_a$,
while
the transfer matrix acts in the quantum space $\otimes_{k=1}^MV_k$.
In  the framework of the quantum inverse scattering method (QISM)
\cite{16,12}
instead of the original non-linear problem the auxiliary linear one is
considered:
\begin{equation}
\label{34}
\frac{d}{dx}T(\lambda,x)=L(\lambda,x)T(\lambda,x)
\end{equation}
or
$$T(n+1,\lambda)=L_{n+1}(\lambda)T(n,\lambda)$$
in discrete case.

This is the Lax operator of the QISM \cite{17}. The solution of (\ref{34}):
\begin{equation} 
\label{35}
T(\lambda,x)=P\exp(\int^xL(\lambda,y)dy),
\end{equation}
$$T(n,\lambda)=L_{n}(\lambda)L_{n-1}(\lambda)...L_1(\lambda)$$
in discrete case, defines the monodromy matrix $T(\lambda)$. 

Its entries are  the new
variables
(the quantum scattering data), which commutation relations are defined by
\begin{eqnarray}
\label{36}
\sum_{j_1,j_2=1}^{n}
R_{i_1i_2,j_1j_2}(\lambda-\mu)
T_{j_1k_1}(\lambda)
T_{j_2k_2}(\mu)=\\
\sum_{j_1,j_2=1}^{n}
T_{i_2j_2}(\mu)
T_{i_1j_1}(\lambda)
R_{j_1j_2,k_1k_2}(\lambda-\mu).\nonumber
\end{eqnarray}
We see that integrable systems are specified by the $R$-matrix, which
acts on $\IC^n\otimes\IC^n$ and satisfies the Yang-Baxter equation
\begin{eqnarray}
\label{37}
\sum_{j_1,j_2,j_3=1}^{n}
R_{i_1i_2,j_1j_2}(\lambda)
R_{j_1i_3,k_1j_3}(\lambda+\mu)
R_{j_2j_3,k_2k_3}(\mu)=\\
\sum_{j_1,j_2,j_3=1}^{n}
R_{i_2i_3,j_2j_3}(\mu)
R_{i_1j_3,j_1k_3}(\lambda+\mu)
R_{j_1j_2,k_1k_2}(\lambda).\nonumber
\end{eqnarray}
In general the $R$-matrix depends on the spectral parameter $\lambda$ and
other parametres. Although there is no complete mathematical theory of the
Yang-Baxter equation, a variety of solutions are! known as well as
different
fields of their application. They are classified by the Lie algebra, its
irreducible representations, and the spectral parameter dependence:
rational, trigonometric and elliptic ones \cite{15}.
Given a solution $R(\lambda)$ one can define the quadratic algebra ${\cal T}_R$
of $n\times n$ matrix elements $T_{ij}$, which is generated by eq.
(\ref{36}).
The associative algebra ${\cal T}_R$ realizes the representation space
of a quantum integrable system.  The commutative integrals of motion
are $t(\lambda)={\rm tr} T(\lambda)$, which follows   from (\ref{36}),
taking the trace of $T_1T_2=R^{-1}_{12}T_2T_1R_{12}$.
The algebra ${\cal T}_R$ posesses the comultiplication property: if
$T_1(\lambda)$ and $T_2(\lambda)$ are two representations of ${\cal T}_R$
in the quantum spaces $V_1$ and $V_2$, then the matrix
\begin{equation}
\label{38}
T_{ik}(\lambda)=T_{1,ij}(\lambda)T_{2,jk}(\lambda)
\end{equation}
is a representation of ${\cal T}_R$ in the tensor product $V_1\otimes V_2$.
This property allows to represent $T(\lambda)$ as a product of
elementary
representations, the so called Lax operators $L_i(\lambda)$. It follows from
$SL(2)$
symmetry of the $R$-matrix that an arbitrary constant $d\times d$ matrix $K$
provides the simplest representation of the algebra ${\cal T}_R$. This
algebra has a central element, the quantum determinant of $T(\lambda)$:
\begin{eqnarray}
\label{39}
\Delta(\lambda)\equiv{\det}_q T(\lambda)=
D(\lambda+\eta/2)A(\lambda-\eta/2)-B(\lambda-\eta/2)C(\lambda+\eta/2=
\\\nonumber
A(\lambda-\eta/2)D(\lambda+\eta/2)-C(\lambda-\eta/2)C(\lambda+\eta/2)=
\\\nonumber
A(\lambda+\eta/2)D(\lambda-\eta/2)-B(\lambda+\eta/2)C(\lambda-\eta/2)=
\\\nonumber
D(\lambda+\eta/2)A(\lambda-\eta/2)-C(\lambda+\eta/2)B(\lambda-\eta/2)=
\nonumber
\end{eqnarray}
which has the following remarkable properties
\begin{equation}
\label{40}
{\det}_qT_1(\lambda)T_2(\lambda)={\det}_q T_1(\lambda){\det}_qT_2(\lambda)
\end{equation}
and
\begin{equation}
\label{41}
{\det}_q K=\det K.
\end{equation}
The next representation is given by Lax operator, mentioned above, which
takes the especially simple form for the $XXX$ spin chain:
\begin{equation}
\label{42}
L(\lambda)=\lambda+\eta\sum_{\alpha=1}^3S_\alpha\sigma_\alpha=\left(\matrix{
\lambda+\eta S^0&\eta S^-\cr-\eta S^+&\lambda-\eta S^0\cr}\right)
\end{equation}
where operators $S_\alpha$ belonging to some irreducible representaion of
$sl(2)$ have commutation relation (\ref{22}). Note that the $R$-matrix
itself
can be choosen as a Lax operator, if the  auxiliary space is
two-dimensional. We have
\begin{equation}
\label{43}
{\det}_q L(\lambda)=\lambda^2-\eta^2({\cal C}+1/4), \ \
{\cal C} = (S^0)^2 - \frac{1}{2} (S^+ S^- + S^- S^+ ).
\end{equation}
Since the $L(\lambda)$-operator, being the elementary representation of
${\cal T}_R$, satisfies the Yang-Baxter relation and the $R$-matrix 
depends only on
the difference of the spectral parameters, the  shift
$L(\lambda)\rightarrow
L(\lambda-\omega)$ defines an automorphysm in ${\cal T}_R$:
\begin{equation}
\label{44}
R (\lambda-\omega_1+S_1^\alpha\sigma_\alpha) \
(\lambda-\omega_2+S_2^\beta\sigma_\beta)=
(\lambda-\omega_2+S_2^\beta\sigma_\beta)
(\lambda-\omega_1+S_1^\alpha\sigma_\alpha) R
\end{equation}
Separating the terms, linear in $\lambda$ in this  equation one deduces
that the $R$-matrix is $SL(2)$-invariant
\begin{equation}
\label{45}
[R;S_1^\alpha+S_2^\alpha]=0
\end{equation}
and depends only on difference $\omega_{12}=\omega_1-\omega_2$.
The $SL(2)$-invariance implies that the  $R$-matrix has to have the
form
\begin{equation}
\label{46}
R=\sum\rho_j(\omega_{12})P_j,
\end{equation}
where $P_j$ are the projectors  corresponding to the decomposition of
the tensor
product of two initial representations into the sum of irreducible
representations labelled by spin $j$.

Furthermore the part of eq.(\ref{44}), which contains no $\lambda$ gives
for $\rho_j(\omega_{12})$ the recurrence relation:
\begin{equation}
\label{47}
\rho_{j+1}(\omega_{12})=\frac
{\omega_{12}+\eta(j+1)}
{\omega_{12}-\eta(j+1)}\rho_j(\omega_{12}),
\end{equation}
which determines $R$ up to a scalar factor.

Particular solutions of the Y-B equation have properties, which are
important
for different applications, but which are not necessarily valid for a
given solution: \\
regularity\\ $$\quad R(0)=\rho(0)^{1/2}{\cal P}_{12}$$
P-symmetry\\$$\quad {\cal P}R_{12}(\lambda){\cal P}\equiv R_{21}
(\lambda)=R_{12}(\lambda)$$
T-symmetry\\ $$\quad R_{12}^{t_1t_2}(\lambda)=R_{12}(\lambda)$$
unitarity\\ $$\quad R_{12}(\lambda)=R_{21}(-\lambda)=\rho(\lambda)I$$
crossing symmetry\\
$$\quad R_{12}(\lambda)=V_{(1)}R_{12}^{t_2}(-\lambda-\eta)V_{(1)}^{-1}$$
quasiclassical property\\
$$\quad R(\lambda,\eta)=I+\eta r(\lambda)+{\cal O}(\eta^2),$$
Here the  superscript $t$ denotes matrix transposition, $ r(\lambda)$ is
the
classical $R$-matrix, $\rho(\lambda)$ is an  even scalar function,
$\eta$ is the crossing parameter and $V$ determines the crossing matrix
$M\equiv V^tV=M^t$. The quasiclassical property gives rise to the
direct connection of the quantum model to the corresponding classical one.
Many $R$-matrices have only the combined $PT$-symmetry:
$R_{12}^t(\lambda)=R_{21}(\lambda)$. The regularity is used to extract 
from $t(\lambda)$ the local integrals of motion.

Thus the  general solution of (\ref{44}) is given by
\begin{equation}
\label{48}
T(\lambda,\vec{\omega})=KL_N(\lambda-\omega_N)...L_2(\lambda-\omega_2)L_1(
\lambda-\omega_1)=\left(\matrix{A(\lambda)&B(\lambda)\cr C(\lambda)&
D(\lambda)\cr}\right).
\end{equation}
because any permutation of the multipliers gives the equivalent result
in the algebra
${\cal T}_R$.  Notice, that the $L$-operator is noting else than a
$R$-matrix, acting
in auxiliary and quantum spaces $\IC^2\otimes V_i$: $L_i(\lambda)\equiv
R_{ai}(\lambda)$.

The corresponding quantum determinant is
\begin{equation}
\label{49}
\Delta(\lambda)={\det}_qT(\lambda)=\det K\prod_{i=1}^N((\lambda-\omega_i)^2-
\eta^2({\cal C}_i+1/4))
\end{equation}
Now, it follows from
\begin{equation}
\label{50}
R_{12}(\lambda-\mu)T^{(1)}(\lambda,\vec{\omega})T^{(2)}(\mu,\vec{\omega})=
T^{(2)}(\mu,\vec{\omega})T^{(1)}(\lambda,\vec{\omega})R_{12}(\lambda-\mu)
\end{equation}
that
\begin{equation}
\label{51}
[t(\lambda,\vec{\omega});t(\lambda,\vec{\omega})]=0,
\end{equation}
where $t(\lambda,\vec{\omega})={\rm tr}T(\lambda,\vec{\omega})$,
The trace is taken over the auxiliary space.

Among the integrals of motion (\ref{51}) we look for
 local ones,
i.e.  quantities $H^{(k)}$ $k=1,2,3,..$, which can be expressed as
the sum of local operators,
\begin{equation}
\label{52}
H^{(k)}=\sum_{i=1}^NH^{(k)}_{i,i-1,...,i-k+1}
\end{equation}
The periodicity, $N+1\equiv 1$, is supposed. The local densities
$H^{(k)}_{i,i-1,...,i-k+1}$ should involve only $k$ adjancent spins $S_i, S_{
i+1},..., S_{i-k+1}$.
An important case when such local integrals exist is that of the
homogeneous spin chain, corresponding to equal spins $\Delta_i=\Delta$ and
zero shifts $\vec{\omega}=0$. It has the important  property of
translational
invariance. The corresponding $R$-matrix is regular. The  similarity
transformation
\begin{equation}
\label{53}
US^\alpha_iU^{-1}=S^\alpha_{i+1},\qquad\qquad US_NU^{-1}={\cal K}_1S_1
{\cal K}_1^{-1},
\end{equation}
where ${\cal K}$ permutes the boundary matrix $K$ and with the Lax
operator  $L_1$:
$KL_1 (\lambda)={\cal K}^{-1}L(\lambda)K{\cal K}$. This transformation
generalizes the ordinary translation for the periodic chain ($K=1$) to the
twisted periodic boundary condition, specified by the matrix $K$ and
$U^N\ne1$ in contrast to the case $K=1$, when operator $U$ takes the
especially simple form: $U={\cal P}_{12}{\cal P}_{23}...{\cal P}_{N-1N}$.
The unitarity of $U$ allows to represent it in exponential form
\begin{equation}
\label{54}
U=e^{iP},
\end{equation}
where operator $P$ has the physical meaning of the total momentum of the
chain. The hamiltonian of the model then acquires the form
\begin{equation}
\label{55}
H=\frac{d}{d\lambda}t(\lambda)|_{\lambda=0}=\sum_{i=1}^N\frac{d}
{d\lambda}{\cal P}_{i, i+1}R_{i, i+1}(\lambda)|_{\lambda=0}.
\end{equation}
Faddeev and Korchemsky have shown \cite{9} that the $N$-reggeon
Hamiltonian (\ref{15}),
corresponding to the homogeneous chain can be obtained in this manner.
Unfortunately, the $R$-matrix, corresponding to the inhomogeneous chain
(see below) posesses no regularity property and the corresponding 
Hamiltonian cannot be related to the  derivative of the transfer-matrix 
in a simple way.

The analysis of the integrable systems is modified upon imposing
boundary
conditions different from the  periodic ones. This case is related to
the factorizable
scattering of particles with internal degrees of freedom on a half-line
\cite{18}. The algebraic description involves  additionally a boundary
operator $B$ into the ZF-algebra \cite{19}:
\begin{equation}
\label{56}
Z_a(\lambda)B=K_{ab}(\lambda)Z_b(-\lambda)B
\end{equation}
Then the two particles factorizability gives rise the reflection equation
(compare (\ref{29}):
\begin{equation}
\label{57}
S_{12}(\lambda-\mu)K_1(\lambda)S_{21}(\lambda+\mu)K_2(\mu)=
K_2(\mu)S_{12}(\lambda+\mu)K_1(\lambda)S_{21}(\lambda-\mu)
\end{equation}
in addition to the Yang-Baxter equation (\ref{30}). The reflection
matrix has
the same properties as the $R$-matrix, regularity: $K(0)=I$; unitarity:
$K(\lambda)K(-\lambda)=I$, T-symmetry: $K^t(\lambda)=K(\lambda)$; the crossing
symmetry is more elaborated and it involves the $S$-matrix as well
\cite{19}.

Then the boundary operator $B$ can be constructed by
\begin{equation}
\label{58}
B=\exp(\int\phi(\lambda)d\lambda)
\end{equation}
from the combination,
\begin{equation}
\label{59}
\phi(\lambda)=Z_a(-\lambda)K_{ab}(\lambda)Z_b(-\lambda),
\end{equation}
which is a "local" field $\phi(\lambda)$: $[\phi(\lambda);\phi(\mu)]=0$.
Due to the  $SL(2)$-symmetry of the $S$-matrix, the corresponding
$K$-matrix
can be transformed $K\rightarrow K^\prime=GKG^{-1}$ with arbitrary $G$
and the general solution of the
reflection equation (\ref{57}) for the rational case, which we are 
interested in, is
\begin{equation}
\label{60}
K(\lambda)=\xi I+\lambda E,\qquad\qquad E^2=I.
\end{equation}
The reflection equation has an important covariance property: if
$T(\lambda)$ and $K(\lambda)$ satisfy the relations (\ref{37}) and
(\ref{57}) then $K^\prime(\lambda)=T(\lambda)K(\lambda)T(-\lambda)^{-1}$
is also a solution of (\ref{79}), provided the entries of $K(\lambda)$
and $T(\lambda)$ commute, $[K_{ab}(\lambda),T_{cd}(\lambda)]=0$.
The proof follows easily by the substitution of $K^{\prime}(\lambda)$
into (\ref{79}) and by using the fundamental Y.B. relation in the
different form $$T_{(2)}^{-1}(-\mu)R_{12}(\lambda+\mu)T_{(1)}(\lambda)=
T_{(1)}^{-1}(\lambda)R_{12}(\lambda+\mu)T_{(2)}(-\mu).$$

If the matrix $T(\lambda)$ is constructed as an ordered product of
$N$ independent Lax operators, then $K^\prime(\lambda)$ can be
interpreted as the monodromy matrix of $N$ site lattice model with
a boundary interaction described by the operator valued entries of
the matrix $K(\lambda)$. It is  called Sklyanin's monodromy matrix.
The corresponding transfer matrix is defined as the trace
$$t(\lambda)={\rm tr}\bar K(\lambda)T(\lambda)K(\lambda)T^{-1}
(-\lambda), $$ where the matrix $\bar K(\lambda)$ is any solution of
(\ref{57}), corresponding to the other boundary,  is commutative
\cite{12} $$[t(\lambda),t(\mu)]=0. $$

In the context of the Heisenberg chain equation (\ref{57}) takes
the form:
\begin{eqnarray}
\label{61,62}
R_{12}(\lambda-\mu)K_{(1)}^-(\lambda)
R_{12}^{t_1t_2}(\lambda+\mu)K_{(2)}^-(\mu)=\\\nonumber
K_{(2)}^-(\mu)R_{12}(\lambda+\mu)K_{(1)}^-(\lambda)
R_{12}^{t_1t_2}(\lambda-\mu),
\\
R_{12}(-\lambda+\mu)(K_{(1)}^+)^{t_1}(\lambda)M_{(1)}^{-1}
R_{12}^{t_1t_2}(-\lambda-\mu-2\eta)M_{(1)}(K_{(2)}^+)^{t_2}(\mu)=\\\nonumber
(K_{(2)}^+)^{t_2}(\mu)M_{(1)}R_{12}(-\lambda-\mu-2\eta)M_{(1)}^{-1}
(K_{(1)}^+)^{t_1}(\lambda)R_{12}^{t_1t_2}(-\lambda+\mu)
\end{eqnarray}
where $M$ is crossing matrix, defined above.
In practice, if $K^{-}(\lambda)$ is a solution of (61) then
$K^+(\lambda)= (K^-(-\lambda-\eta))^tM$ is a solution of (62).
The eq. (\ref{57}) has an important covariance property: if
$T(\lambda,\vec{
\omega})$ and $K_\pm(\lambda)$ satisfies the relations (\ref{37}) and
(61), (62) then Sklyanin's monodromy matrix:
\begin{equation}
\label{63}
U(\lambda,\vec{\omega})=T(\lambda,\vec{\omega})K^-(\lambda)\tilde
T(\lambda,\vec{\omega}),
\end{equation}
where $\tilde T(\lambda,\vec{\omega})=R_{Na}(\lambda-\omega_N)...
R_{2a}(\lambda-\omega_2)R_{1a}(\lambda-\omega_1)$, (cr. with (\ref{58})),
satisfies the relation
\begin{equation}
\label{64}
R_{12}(\lambda-\mu)U_{(1)}(\lambda,\vec{\omega})
R_{12}^{t_1t_2}(\lambda+\mu)U_{(2)}(\mu,\vec{\omega})=
U_{(2)}(\mu,\vec{\omega})R_{12}(\lambda+\mu)U_{(1)}(\lambda,
\vec{\omega})R_{12}^{t_1t_2}(\lambda-\mu).
\end{equation}
Indeed, we note that unitarity and crossing symmetry together
imply the relation
\begin{equation}
\label{65}
M_{(1)}R_{12}^{t_2}(-\lambda-\eta)M_{(1)}^{-1}
R_{12}^{t_2}(\lambda-\eta)=\rho(\lambda).
\end{equation}
Futhermore, we see that unitarity implies $T(\lambda,\vec{\omega})
\tilde T(-\lambda,\vec{\omega})=\prod\rho(\lambda-\omega_i)$.
Therefore, up to a scalar factor, $\tilde T(-\lambda,\vec{\omega})$
is the inverse of $T(\lambda,\vec{\omega})$..

The commutativity of the transfer matrix $t(\lambda,\vec{\omega})$
implies
integrability of the open quantum spin chain with the hamiltonian
\cite{12}:
\begin{equation}
\label{66}
H=\sum^{N-1}_{i=1}H_{ii+1}+1/2(K_-^{(1)})^t+\frac{{\rm tr}_0K_+^{(0)}
(0)H_{N0}}{{\rm tr}K_+(0)}
\end{equation}
whose two-site terms are given by
\begin{equation}
\label{67}
H_{ii+1}=\frac{d}{d\lambda}{\cal P}_{ii+1}R_{ii+1}(\lambda)|_{\lambda=0}
\end{equation}
in the standard fashion.

\section{Closed XXX Heisenberg chain}
There are two ways of including fermions.
The first corresponds to considering closed Heisenberg
chains with different spins, i.e. contain the operators ${\cal G}_{FG}$
and $H_F$ together with $H_G$. This case arises in amplitudes with the
exchange of two adjoint fermions and one gluon. The same hamiltonian
also describes the exchange of three fermions in the fundamental
representation of $SU(3)$ if not all three have the same helicity.
In the first case the Regge singularity is near $j = 0$ and in the
second near $j = -1/2$. The seconds way of including fermions will be
considered in the next section.

We consider the conformally covariant operator obtained from 
$H_F^{\omega}$ by substituting $\omega = 0$,
\begin{equation}
\label{68}
H^{0\frac{1}{2}}=\partial_1^{-1}\log x_{12}\partial_1+\log x_{12}+
\log\partial_1\partial_2 - 2\psi(1)=x_{12}\log\partial_1x_{12}^{-1}+
\log\partial_2+2\log x_{12}-2\psi(1)
\end{equation}
with $\Delta_1=0$, $\Delta_2=1/2$ and the  conjugated operator
\begin{equation}
\label{69}
H^{\frac{1}{2}0}=\partial_2^{-1}\log x_{12}\partial_2+\log x_{12}+
\log\partial_1\partial_2 - 2\psi(1)=x_{12}\log\partial_2x_{12}^{-1}+
\log\partial_1+2\log x_{12}-2\psi(1)
\end{equation}
with $\Delta_1=1/2$, $\Delta_2=0$.
We have also $\tilde H_F$ which should be denoted by
\begin{equation}
\label{70}
H^{\frac{1}{2}\frac{1}{2}}=2\log x_{12}+\log\partial_1\partial_2=
x_{12}^{-1}H^{00}x_{12}
\end{equation}
with the weights $\Delta_1 = \Delta_2 = \frac{1}{2}$.
We have used
the identity
$$(x_{12}\partial_2)^{-1}=\partial_2^{-1}\log x_{12}\partial_2-
\log x_{12}$$
The operator $H^{\frac{1}{2}\frac{1}{2}}$
is selfconjugated.
\begin{equation}
\label{71}
(H^{\frac{1}{2}\frac{1}{2}})^T=
H^{\frac{1}{2}\frac{1}{2}}
\end{equation}
and for $H^{0\frac{1}{2}}$ we have
\begin{eqnarray}
\label{72}
(H^{0\frac{1}{2}})^T
=\partial_1\log x_{12}\partial_1^{-1}+
\log x_{12}+\log\partial_1\partial_2-2\psi(1)=
\partial_1H^{0\frac{1}{2}}\partial_1^{-1}=\\\nonumber
=x_{12}^{-1}\log\partial_1x_{12}+
\log\partial_2+2\log x_{12}-2\psi(1)=
{\cal P}_{12}x_{12}^{-1}H^{0\frac{1}{2}}x_{12}{\cal P}_{12}\nonumber
\end{eqnarray}
Therefore,
\begin{equation}
\label{73}
[H^{0\frac{1}{2}};
{\cal P}_{12}x_{12}\partial_1]=0
\end{equation}
Taking into account that
\begin{equation}
\label{74}
({\cal P}_{12}x_{12}\partial_1)^2=
{\cal P}_{12}x_{12}\partial_1{\cal P}_{12}x_{12}\partial_1=
x_{21}\partial_2x_{12}\partial_1=-{\cal C}^{0\frac{1}{2}}
\end{equation}
and comparing with the general expression
\begin{equation}
\label{75}
{\cal C}^{\Delta_1\Delta_2}=(\vec{S_1}+\vec{S_2})^2=
x_{12}^2\partial_1\partial_2+2x_{12}(\Delta_1\partial_2-\Delta_2
\partial_1)+(\Delta_1+\Delta_2)(1-\Delta_1-\Delta_2)
\end{equation}
we can conclude that indeed under conformal transformations
the operator
$H^{0\frac{1}{2}}$ transforms
covariantly with weights $0, \frac{1}{2}$.\\
Consider first the homogeneous \\
closed chain, which consists\\
out of three fermions.
\qquad\qquad
\parbox{15cm}{
\unitlength=1mm
\linethickness{0.2mm}
\begin{picture}(30,20)
\put(5,5){\circle*{3}}
\put(15,22){\circle*{3}}
\put(25,5){\circle*{3}}
\put(6.5,5){\line(1,0){17}}
\put(5,6.5){\line(3,5){8.6}}
\put(25,6.5){\line(-3,5){8.6}}
\put(2,7){\shortstack{1}}
\put(14,24){\shortstack{2}}
\put(26,7){\shortstack{3}}
\put(10,0){\shortstack{Fig.2}}
\end{picture}
}
The corresponding hamiltonian is
\begin{eqnarray}
\label{76}
H_{123}^{\frac{1}{2}\frac{1}{2}\frac{1}{2}}=H_{32}^{\frac{1}{2}}+
H_{21}^{\frac{1}{2}\frac{1}{2}}+H_{13}^{\frac{1}{2}\frac{1}{2}}=\\
2\log x_{12}x_{23}x_{31}+2\log\partial_1\partial_2\partial_3.\nonumber
\end{eqnarray}
Conjugation arguments as described above give nothing  for
$H^{\frac{1}{2}\frac{1}{2}}$, while the commutators of
particular Casimir operators provide us with two conserved currents:
\begin{eqnarray}
\label{77}
D^{(3)}_{\frac{1}{2}\frac{1}{2}\frac{1}{2}}=x_{12}x_{23}x_{31}\partial_1
\partial_2\partial_3+1/2(x_{23}(x_{31}-x_{12})\partial_2\partial_3+\\
\nonumber
x_{12}(x_{23}-x_{31})\partial_1\partial_2+x_{31}(x_{12}-x_{23})\partial_3
\partial_1)-1/2(x_{12}\partial_3+x_{23}\partial_1+x_{31}\partial_2).
\end{eqnarray}
is a third order differential operators and
\begin{eqnarray}
D^{(2)}_{\frac{1}{2}\frac{1}{2}\frac{1}{2}} =
(x_{12}^2\partial_1\partial_2+x_{23}^2\partial_2
\partial_3+x_{31}^2\partial_3\partial_1-x_{31}\partial_3+x_{12}
\partial_2)+\\
\nonumber
(x_{31}-x_{12})\partial_1+(x_{12}-x_{23})\partial_2
+(x_{23}-x_{31})\partial_3,
\end{eqnarray}
is a  second order one. The latter reflects the conformal
symmetry of the system. The relations
\begin{equation}
\label{78}
[H_{123}^{\frac{1}{2}\frac{1}{2}\frac{1}{2}},
D^{(2)}_{\frac{1}{2}\frac{1}{2}\frac{1}{2}}]=0
\end{equation}
and
\begin{equation}
\label{79}
[H_{123}^{\frac{1}{2}\frac{1}{2}\frac{1}{2}},
D^{(3)}_{\frac{1}{2}\frac{1}{2}\frac{1}{2}}]=0
\end{equation}
can be checked by direct calculations.
These conserved currents can be obtained also in more regular way. They
appear as coefficients in front of $\lambda^{1}$
and $\lambda^{0}$ in the monodromy matrix expansion $t(\lambda)$,
where
\begin{equation}
\label{80}
t(\lambda)=tr(L_1^{1/2}(\lambda)L_2^{1/2}(\lambda)L_3^{1/2}(\lambda))=
D^{(3)}_{\frac{1}{2}\frac{1}{2}\frac{1}{2}}+\lambda
D^{(2)}_{\frac{1}{2}\frac{1}{2}\frac{1}{2}}+\lambda^3+1/4
\end{equation}
and the Lax operators $L_i^\Delta(\lambda)$ are defined in (\ref{23})
with $\Delta_i=1/2$.

Let us consider now\\ the inhomogeneous closed chain,\\ which
consisits of\\ one fermion and two gluons.
\qquad\qquad
\parbox{15cm}{
\unitlength=1mm
\linethickness{0.2mm}
\begin{picture}(30,20)
\put(5,5){\circle*{3}}
\put(15,22){\circle{3}}
\put(25,5){\circle{3}}
\put(6.5,5){\line(1,0){17}}
\put(5,6.5){\line(3,5){8.6}}
\put(25,6.5){\line(-3,5){8.6}}
\put(2,7){\shortstack{1}}
\put(14,24){\shortstack{2}}
\put(26,7){\shortstack{3}}
\put(10,0){\shortstack{Fig.3}}
\end{picture}}
\
The corresponding hamiltonian is
\begin{eqnarray}
\label{81}
\tilde H_{123}^{\frac{1}{2}00}=
H_{32}^{00}+H_{21}^{0\frac{1}{2}}+H_{13}^{\frac{1}{2}0}=\\
2\log x_{12}x_{23}x_{31}+2\log\partial_1\partial_2\partial_3+\partial_3^
{-1}(x_{23}^{-1}-x_{31}^{-1})+\partial_2^{-1}(x_{12}^{-1}-x_{23}^{-1}).
\nonumber
\end{eqnarray}
Transposition arguments do not work here. However,
for closed chains with three sites there is always a third order
differential
operator, commuting with operator of total spin of chain ${\cal C}$.
Indeed, the result for the commutator
\begin{eqnarray}
\label{82}
[{\cal C}^{\Delta_1\Delta_2}_{12}; {\cal C}^{\Delta_3\Delta_1}]=
[x_{12}^2\partial_1\partial_2+2x_{12}(\Delta_1\partial_2-\Delta_2
\partial_1);x_{31}^2\partial_3\partial_1+2x_{31}(\Delta_3\partial_1-
\Delta_1\partial_3)]=\\\nonumber
x_{12}x_{23}x_{31}\partial_1\partial_2\partial_3+\Delta_3x_{12}(x_{23}-
x_{31})\partial_1\partial_2+\Delta_2x_{31}(x_{12}-x_{23})\partial_3
\partial_1+\Delta_1x_{23}(x_{31}-x_{12})\partial_2\partial_3\\\nonumber
-2(\Delta
_2\Delta_3x_{23}\partial_1+\Delta_3\Delta_1x_{31}\partial_2+\Delta_1
\Delta_2x_{12}\partial_3)\equiv D^{(3)}_{\Delta_1\Delta_2\Delta_3}
\end{eqnarray}
is symmetric under cyclic permutation of $(123)$. Therefore its
commutator
with $D^{(2)}_{\Delta_1\Delta_2\Delta_3}\equiv {\cal C}^{\Delta_1\Delta_2}
_{12}+{\cal C}^{\Delta_2\Delta_3}_{23}+{\cal C}^{\Delta_3\Delta_1}_{31}$
vanishes due to the Jacoby identity.
However, the hamiltonian $\tilde H_{123}^{\frac{1}{2}00}$
commuting with $D^{(2)}_{\frac{1}{2}00}$ does not commute with
$D^{(3)}_{\frac{1}{2}00}$:
\begin{eqnarray}
\label{83}
[ \tilde H_{123}, D^{(3)} ] = \\ \nonumber
[(2\log x_{12}x_{23}x_{31}+2\log\partial_1\partial_2\partial_3+
\partial_3^{-1}(x_{23}^{-1}-x_{31}^{-1})+\partial_2^{-1}(x_{12}^
{-1}-x_{23}^{-1})),\\(x_{12}x_{23}x_{31}\partial_1\partial_2\partial_3+
1/2(x^2_{12}-x^2_{31})\partial_2\partial_3)]=
\frac{1}{2} \partial^{-1}_3 x^2_{12} x^{-2}_{31}\partial_2
-\frac{1}{2} \partial^{-1}_2 x^2_{31} x^{-2}_{12}\partial_3.
\nonumber
\end{eqnarray}

These operators $D^{(2)}, D^{(3)}$ also appear as coefficients in front
of $\lambda^{1}$
and $\lambda^{0}$ in the monodromy matrix expansion $t(\lambda)$, where
\begin{equation}
\label{84}
t(\lambda)=tr(L_1^{1/2}(\lambda)L_2^0(\lambda)L_3^0(\lambda))=
D^{(3)}+\lambda D^{(2)}+\lambda^2(2\lambda-1).
\end{equation}
The next inhomogeneous chain is\\ the one with one gluon\\ and two fermions:
\qquad\qquad
\parbox{15cm}{
\unitlength=1mm
\linethickness{0.2mm}
\begin{picture}(30,20)
\put(5,5){\circle*{3}}
\put(15,22){\circle{3}}
\put(25,5){\circle*{3}}
\put(6.5,5){\line(1,0){17}}
\put(5,6.5){\line(3,5){8.6}}
\put(25,6.5){\line(-3,5){8.6}}
\put(2,7){\shortstack{1}}
\put(14,24){\shortstack{2}}
\put(26,7){\shortstack{3}}
\put(10,0){\shortstack{Fig.4}}
\end{picture}}
\
The corresponding hamiltonian commutes with the Casimir operator
\begin{equation}
\label{85}
[(x_{12}\partial_1x_{12}\partial_2+x_{23}\partial_3x_{23}\partial_2+
x_{31}\partial_1\partial_3x_{31});(2\log x_{12}x_{23}x_{31}+
2\log\partial_1\partial_2\partial_3+
\partial_2^{-1}(x_{12}^{-1}-x_{23}^{-1})]=0,
\end{equation}
and does not commute with the next current
\begin{eqnarray}
\label{86}
\hspace{-1cm}
[(x_{23}\partial_3x_{31}\partial_1x_{12}\partial_2-\frac{1}{2}
x_{31}\partial_1\partial_3x_{31});(2\log x_{12}x_{23}x_{31}+
2\log\partial_1\partial_2\partial_3+
\partial_2^{-1}(x_{12}^{-1}-x_{23}^{-1})]=\nonumber\\
\frac{1}{2}\partial_2^{-1}x_{31}(x_{23}^{-2}\partial_1-x_{12}^{-2}
\partial_3)x_{31}.
\end{eqnarray}

The same happens for the longer closed chains.
The hamiltonians for that chains commute 
only with the first current, the Casimir operator.
This means that in order to describe the integrable model, the
corresponding hamiltonian should be modified. Indeed, the 
inhomogeneous chains (fig. 3,4) have to be considered as the 
chains with impurity. Probably, the expression for
$H^{0\frac{1}{2}}$ has to be changed slightly.

\section{XXX Heisenberg chain with open boundary}
The second way of including fermions is to build an open chain with the
fermions (now in the fundamental gauge group representation) at the
ends. This open chain corresponds
to the Regge exchange with meson quantum numbers in the $t$-channel.

Let us consider the hamiltonian
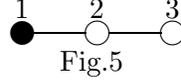
\renewcommand{\baselinestretch}{1.2}
\qquad
\parbox{7cm}{
\unitlength=1mm
\linethickness{0.2mm}
\begin{picture}(30,20)
\put(5,5){\circle*{3}}
\put(15,5){\circle{3}}
\put(25,5){\circle{3}}
\put(6.5,5){\line(1,0){7}}
\put(16.5,5){\line(1,0){7}}
\put(4,7){\shortstack{1}}
\put(14,7){\shortstack{2}}
\put(24,7){\shortstack{3}}
\put(10,0){\shortstack{Fig.5}}
\end{picture}}
\quad
corresponding to the following chain

\begin{eqnarray}
\label{87}
 H_{123}^{\frac{1}{2}00}=H_{32}^{00}+H_{21}^{0\frac{1}{2}}=\nonumber\\
\log x_{12}+\partial_2^{-1}\log x_{32}x_{21}\partial_2+\log\partial_1
\partial_2^2\partial_3+\partial_3^{-1}\log x_{32}\partial_3=\nonumber\\
\log\partial_1+x_{32}\log\partial_2\partial_3x_{32}^{-1}+
x_{21}\log\partial_2x_{21}^{-1}+2\log x_{21}x_{32},\nonumber
\end{eqnarray}
The transposed Hamiltonian has form
\begin{equation}
\label{88}
( H^{\frac{1}{2}00}_{123})^T=
\partial_2\partial_3 H_{123}\partial_2\partial_3=
{\cal P}_{123}(x_{32}x_{21})^{-1} H_{123}x_{32}x_{21}{\cal P}_{123}.
\end{equation}
Then $ H$ commutes with
\begin{equation}
\label{89}
A_{123}^{\frac{1}{2}00}={\cal P}_{123}x_{32}x_{21}
\partial_2\partial_3{\cal P}_{123}.
\end{equation}
The permutation operator ${\cal P}_{123}$ maps the sites 1,2,3 of the
chain into the sites 3,2,1 correspondingly.

In a  similar way the  conserved operator of the highest
order in the derivatives can be obtained for the open chain with $N-2$
gluonic operators and  fermions of opposite spin at the ends:
\begin{equation}
\label{90}
H_{12...N}^{\frac{1}{2}0...0}=\partial_2^{-1}\log x_{21}\partial_2+
\log x_{21}+\log\partial_2\partial_1+\sum^{N-1}_{i=2}H_{ii+1}^{00}
\end{equation}
commutes with the charge operator
\begin{equation}
\label{91}
A_{12...N}={\cal P}_{12...N}x_{NN-1}...x_{21}\partial_N\partial_
{N-1}...\partial_2.
\end{equation}

For simplicity we put the reflection matrices $K^{\pm}$ equal to unity.
 From eq. (\ref{56})
one can see that this implies the transfer matrix   to be an even
function of the spectral parameter. In our case we obtain up to
insignificant c-number terms
\begin{equation}
\label{92}
t(\lambda)=(4\lambda^2-1)(D^{(4)}-\omega_2D^{(3)}_{\frac{1}{2}00}-
(\lambda^2-\omega_2^2)D^{(2)}_{\frac{1}{2}00}+(\lambda^2-\omega_2^2)^2),
\end{equation}
where $D^{(3)}_{\frac{1}{2}00}$ and
$D^{(2)}_{\frac{1}{2}00}$
are the same as in (\ref{38}) for the closed
chain and $$D_{\frac{1}{2}00}^{(4)}=
(x_{12}^2x_{23}^2\partial_1\partial_2+x_{12}x_{23}
(x_{12}-x_{23})\partial_1+x_{12}x_{23}^2\partial_2+x_{23}(x_{12}-x_{23})
\partial_2\partial_3.$$

This coincides with the square of the operator (\ref{75}), obtained by 
transposition arguments.

We put here $\omega_3^2=\omega_2^2= \omega_1^2+3/4$
in order to achieve that the second order operator $D^{(2)}$ coincides
with ${\cal C}^{01/2}$.
Now the operator $D^{(2)}$ commutes with hamiltonian.
Thus, for the open chain (Fig. 5), described by the hamiltonian $
H_{123}$
the eigenvalue problem like eq. (\ref{67}) can be replaced by the
one with the same Casimir operator
 and with the fourth order differential operator
$D^{(4)}$ instead of $D^{(3)}=\tilde A$ as for the closed chain.

So we see that for open chains we are able to find conserved currents and
its number is just enough to solve the! eigenvalue problem for QCD
reggeon interactions.
Moreover, for the open chain Fig. 5 with one fermion inside again
everything is all right. The hamiltonian
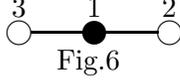
\renewcommand{\baselinestretch}{1.2}
\qquad
\parbox{7cm}{
\unitlength=1mm
\linethickness{0.2mm}
\begin{picture}(30,20)
\put(5,5){\circle{3}}
\put(15,5){\circle*{3}}
\put(25,5){\circle{3}}
\put(6.5,5){\line(1,0){7}}
\put(16.5,5){\line(1,0){7}}
\put(4,7){\shortstack{3}}
\put(14,7){\shortstack{1}}
\put(24,7){\shortstack{2}}
\put(10,0){\shortstack{Fig.6}}
\end{picture}}
and the transfer matrix are given by
\begin{equation}
H^{0\frac{1}{2}0}_{312}=H_{31}^{0\frac{1}{2}}+H_{12}^{\frac{1}{2}0}=
2\log x_{12}x_{31}+2\log\partial_1^2\partial_2\partial_3+\partial^{-1}_2
\log x_{12}-\partial_3^{-1}\log x_{31}
\end{equation}
\begin{equation}
t(\lambda)=4\lambda(\lambda+1)x_{31}\partial_1x_{12}^2\partial_1x_{31}
\partial_2\partial_3+4\lambda^2(\lambda+1)^2(x_{23}\partial_2\partial_3+
x_{31}\partial_1x_{31}\partial_3+x_{12}\partial_1x_{12}\partial_2
\end{equation}

The case of the open chain with two fermions and one gluon can be considered 
as well.\\ 
For the chain 
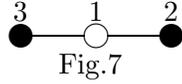
\renewcommand{\baselinestretch}{1.2}
\qquad
\parbox{7cm}{
\unitlength=1mm
\linethickness{0.2mm}
\begin{picture}(30,20)
\put(5,5){\circle*{3}}
\put(15,5){\circle{3}}
\put(25,5){\circle*{3}}
\put(6.5,5){\line(1,0){7}}
\put(16.5,5){\line(1,0){7}}
\put(4,7){\shortstack{3}}
\put(14,7){\shortstack{1}}
\put(24,7){\shortstack{2}}
\put(10,0){\shortstack{Fig.7}}
\end{picture}}
\vspace{0.5cm}
\\
we have two conserved currents:
\begin{eqnarray}
&&[x_{12}x_{23}\partial_1\partial_2\partial_3x_{12}x_{23}\partial_2;
(2\log x_{12}x_{23}+\log\partial_1\partial_2^2\partial_3+
\partial_2^{-1}(x_{12}^{-1}-x_{23}^{-1})]=0\\
&&\hspace{-1cm}
[(x_{12}\partial_1x_{12}\partial_2+x_{23}\partial_3x_{23}\partial_2+
x_{31}\partial_1\partial_3x_{31});
(2\log x_{12}x_{23}+\log\partial_1\partial_2^2\partial_3+
\partial_2^{-1}(x_{12}^{-1}-x_{23}^{-1}))]=0
\end{eqnarray}
Another chain with two fermions and one gluon is:
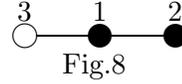
\renewcommand{\baselinestretch}{1.2}
\qquad
\parbox{7cm}{
\unitlength=1mm
\linethickness{0.2mm}
\begin{picture}(30,20)
\put(5,5){\circle{3}}
\put(15,5){\circle*{3}}
\put(25,5){\circle*{3}}
\put(6.5,5){\line(1,0){7}}
\put(16.5,5){\line(1,0){7}}
\put(4,7){\shortstack{3}}
\put(14,7){\shortstack{1}}
\put(24,7){\shortstack{2}}
\put(10,0){\shortstack{Fig.8}}
\end{picture}}
\vspace{0.5cm}
\\
We have following relations for this chain
\begin{eqnarray}
&&[x_{12}\partial_2x_{23}\partial_3x_{23}\partial_2x_{12}\partial_1;
(2\log x_{12}x_{23}+\log\partial_1\partial_2^2\partial_3-
\partial_1^{-1}x_{12}^{-1})]=0\\
&&\hspace{-1cm}
[(x_{12}\partial_2x_{12}\partial_1+x_{23}\partial_2\partial_3x_{23}+
x_{31}\partial_3x_{31}\partial_1);
(2\log x_{12}x_{23}+\log\partial_1\partial_2^2\partial_3-
\partial_1^{-1}(x_{12}^{-1})]=0
\end{eqnarray}

For    completeness \\ let us consider also\\ 
the open chain with three fermions
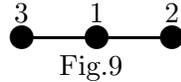
\renewcommand{\baselinestretch}{1.2}
\qquad
\parbox{7cm}{
\unitlength=1mm
\linethickness{0.2mm}
\begin{picture}(30,20)
\put(5,5){\circle*{3}}
\put(15,5){\circle*{3}}
\put(25,5){\circle*{3}}
\put(6.5,5){\line(1,0){7}}
\put(16.5,5){\line(1,0){7}}
\put(4,7){\shortstack{3}}
\put(14,7){\shortstack{1}}
\put(24,7){\shortstack{2}}
\put(10,0){\shortstack{Fig.9}}
\end{picture}}
\vspace{0.5cm}
\\
we can deduce following relations:
\begin{eqnarray}
&&[(\partial_1x_{12}\partial_2x_{23}\partial_3x_{23}\partial_2x_{12}+
x_{12}\partial_2x_{23}\partial_3x_{23}\partial_2x_{12}\partial_1);
(2\log x_{12}x_{23}+\log\partial_1\partial_2^2\partial_3)]=0\\
&&
[(x_{12}\partial_1\partial_2x_{12}+x_{23}\partial_3\partial_2x_{23}+
x_{31}\partial_1\partial_3x_{31});
(2\log x_{12}x_{23}+\log\partial_1\partial_2^2\partial_3)]=0
\end{eqnarray}
Thus,  all  possible open chains with three sites corresponding to 
gluonic or fermionic reggeons are integrable. The conserved currents 
can be derived from the transfer matrix with boundaries represented by  
unity reflection matrices. The hamiltonians are constructed from the 
QCD kernels given above and commute with these two conserved currents 
given by the second and fourth order differential operators respectively. 
We believe  that integrability holds for the longer open chains too.

\section{Conclusions}
The examples considered above, except for the difficulties with the
inhomogeneous closed chains, show the validity of  Lipatov`s
conjecture about the deep connection between the kernels of high-energy
QCD scattering amplitudes and the exactly solvable two-dimensional models
in the cases with fermions as well. In the pure gluonic this connction
context has already been used for the solution of the Odderon problem
\cite{20}. The case of open chain differs from the closed chain case by
the higher order of non-trivial conserved charge; the fourth order appears
instead of the third.

Recently it has been shown that the considered integrable structures
appear also in hard (exclusive and deep-inelastic) scattering \cite{21},
corresponding to a different limiting  region of high energy scattering
amplitudes in QCD.

\end{document}